\documentclass[preprint2]{aastex}
\input epsf

\newcommand{\etal}{{\it et al}}

\slugcomment{Submitted to Astrophys. J. {\it Letters}}

\shorttitle{Fluorescent Excitation of Spectral Lines}
\shortauthors{Chen and Pradhan}

\begin{document}

\title{Fluorescent Excitation of Spectral Lines In Planetary Nebulae}

\author{Guo Xin Chen and Anil K. Pradhan}
\affil{Department of Astronomy, The Ohio State University, Columbus, OH
43210}

\begin{abstract}
   Fluorescent excitation of spectral lines 
is demonstrated as a function of
temperature-luminosity and the distance of the emitting region 
from the central stars of planetary nebulae. The  electron
densities and temperatures are determined, and the method is exemplified 
through a detailed analysis of spectral observations of a high excitation 
PN, NGC 6741, observed by \citet{hyu97}. Fluorescence should
also be important in the determination of element abundances.
It is suggested that the method could be
generally applied to determine or constrain the luminosity
and the region of spectral emission in other intensively radiative
sources such as novae, supernovae, and active galactic nuclei.

\end{abstract}


\keywords{planetary nebulae: general --- H II regions --- atomic data
--- line: formation --- radiation mechanisms: thermal --- radiative
transfer}


\section{INTRODUCTION}

 In a classic paper \citep{sea78} discussed the excitation of nebular
emission lines by resonant scattering of radiation from the central
stars, and found that in addition to electron scattering and recombination,  
absorption of radiation should be taken into account in the analysis of
O~III lines and consequent determination of oxygen abundance. The
high intensity of the continuum radiation from planetary
nebulae, and other sources, are likely candidates for such 
fluorescent excitation (hereafter
FLE) of spectral lines, provided the atomic structure of the given ionic
species (such as O~III) leads itself to resonant excitation from the ground state,
followed by cascades into upper levels of observed transitions. Given
that FLE is operative, it follows that the intensities of spectral lines
(or line ratios) would also be dependent on the radiation
temperature-luminosity of the source and the region of the dominant
abundance of the ionic species, i.e. the distance from the source, via a
geometrical dilution factor. It was shown by \citet{luc95} that the
nickel overabundance problem in a variety of gaseous nebulae may
be addressed by taking into account the FLE of [Ni II] optical lines by the
background stellar UV continuum. Following \citet{luc95}, 
\citet{bau96} and \citet{bau98} investigated FLE in [Fe II] optical transitions, but
found it to be less effective in line formation than [Ni~II] 
owing to differences in atomic structure and associated transition 
probabilities.

 In the present {\it Letter} we explore FLE in a more general manner,
using optical emission lines from a moderately high ionization state 
of iron, Fe~VI, observed in several planetary nebulae in extensive
observational studies by Hyung and Aller (e.g.
1995,1996,1997,1998). In particular we focus on NGC 6741 \citep{hyu97}
where we find clear evidence of FLE. It is shown that the observed
[Fe~VI] line ratios can only be explained with FLE, and yield
electron densities lower than those derived from other less ionized
species such as the well known
[O~II] and [S~II] fine structure doublets \citep{ost89}. The inferred
distances of the [Fe~VI]
emission region from the central stars is consistent with 
the $He^{2+}$ zone (\cite{sea78}), within the inner radius
of the nebulosity. Possible implications of this work are discussed in
relation to other sources, such as novae and AGN, 
where FLE of spectral lines should be important in the determination 
of  densities,temperatures, abundances, and the spatial extent of
emission regions. It is shown that surface 
contour plots of line ratios as double-valued function of radiation
temperature of the source, and the dilution factor, may serve to constrain
or determine both quantities from theoretical calculations and
observations.

\section{FLUORESCENT EXCITATION OF Fe~VI}

 Given an external radiation field, atomic FLE can be effective for 
excitation from the ground state and subsequent cascades to levels 
via dipole allowed transitions. The $\Delta S = 0$ transitions,
where (2S+1) is the spin multiplicity, are generally more efficient than the
intercombination $\Delta S \neq 0$ cascade transitions, since the 
former usually have orders of magnitude higher transition probabilities.
 A schematic representation of the energy level structure of 
Fe~VI with FLE is given in Fig. 1. The dominant UV excitations from the 
even parity ground LS term $3d^3 \ (^4F_J)$ and fine structure levels $J = 3/2,5/2,7/2,
9/2$ to the odd parity levels of the terms of the excited configuration
$3d^2 \ 4p \ (^4G^o_J,^4F^o_J,^4D^o_J)$ are strong dipole allowed
transitions at 3.08 - 3.15 Rydbergs in the far UV range 289 - 296 $\AA$.

 A collisional-radiative model with FLE involving 80 fine structure levels 
for Fe VI is employed to calculate level
populations, $N_i$, relative to the ground level. The emitted flux
per ion for transition $ j\rightarrow i$,
or the emissivity $\epsilon_{ji}$ (ergs$~$cm$^{-3}$s$^{-1}$), is given by,
$\epsilon_{ji}=N_jA_{ji}h\nu _{ij}$. In matrix notation the
coupled equations of statistical equilibrium for the optically thin
case can be expressed as:
{\bf C=N$_e$Q+A+JB}, where $N_e$ is the electron density, {\bf Q}, {\bf A},
and {\bf B} are
 the collisional excitation and the Einstein spontaneous and 
stimulated emission rate matrices, and {\bf J} represents the photon density from a 
continuum radiation field. The matrix elements of the total excitation
rate C$_{ij}$ are then expressed as:

\begin{equation}
\begin{array}{c}
C_{ij}=q_{ij}N_e+J_{ij}B_{ij} ~~~(j>i),\\

C_{ij}=q_{ij}N_e+A_{ij}+J_{ij}B_{ij} ~~~(j<i),
\end{array}
\end{equation}
where $J_{ij}$ are the mean intensities of the continuum at the frequency
for transition $i\rightarrow j$. For example, for a thermal source at 
effective temperature $T_{eff}$, the monochromatic
flux at the photosphere $F_{\nu}$ is given by the Planck function

\begin{equation}
J_{ij}=WF_{\nu}=W\frac{8\pi h\nu ^3}{c^2}\frac 1{e^{h\nu/kT_{eff}}-1}
\end{equation}

where $W=\frac 14(\frac {R_{\ast}}r)^2=1.27\times 10^{-16} (\frac {R_{\ast}/R_{\odot}}{r/pc})^2$
is the geometrical dilution factor; $R_{\ast}$ and r are
the radius of the photosphere and the distance between the star
and the emission region respectively.

 All required atomic data have been recently computed by the
authors under the Iron Project \citep{hum93}. This includes
collision strengths for 3,160 transitions among the 80 Fe~VI levels
\citep{che99a}, and transition probabilities for 867 allowed
electric dipole (E1) and 1,230 forbidden electric quadrupole and
magnetic dipole (E2,M1) transitions \citep{che99c}.
While the E2,M1 A-values agree reasonably well with the earlier work on
[Fe~VI] by \citet{nus78}, the present collision strength data 
differs considerably from theirs owing to resonance effects
(\citet{nus78} did not consider the FLE mechanism for [Fe~VI] line
formation).  A detailed discussion is given by \citet{che99c}, which 
also presents more extensive results on the computed
Fe~VI line emissivity ratios than considered in this {\it Letter}).

\section{PLANETARY NEBULA NGC 6741}

 Observations of this high excitation nebula by \citet{all85} and \citet{hyu97} show
several optical [Fe~VI] lines in the spectrum from the multiplet $3d^3 \ (^4F - ^4P)$
at 5177, 5278, 5336, 5425, 5485, 5632 and 5678 $\AA$ and from
the $(^4F - ^2G)$ at 4973 and 5147 $\AA$. The basic observational
parameters, in particular the inner and the outer radii needed to
estimate the distance from the central star and the dilution factor, 
are described in these works, and their diagnostic diagrams based on the
spectra of a number of
ions give $T_e = 12500K, N_e = 6300 cm^{-3}$, and a stellar $T_{eff} = 
140,000 K$.
As the ionization potentials of Fe~V and Fe~VI are 75.5 eV and 100 eV
respectively, compared to that of He~II at 54.4 eV, 
Fe~VI emission should stem from the fully ionized $He^{2+}$ zone, and
within the inner radius, i.e. r(Fe~VI) $\leq r_{in}$. With these
parameters we obtain the dilution factor to be W = 10$^{-14}$; 
the dominant [Fe~VI] emission region could be up to a factor of 3 closer to
the star, with W up to 10$^{-13}$, without large variations in the
results obtained.

 Fig. 2 presents 4 [Fe~VI] line ratios as a function of several
parameters, in particular with and without FLE. 
In all cases the FLE = 0 curve fails to correlate with the
observed line ratios, and shows no dependence on N$_e$ (an unphysical
result), whereas with FLE we obtain
a consistent $N_e \approx$ 1000-2000 cm$^{-3}$, 
suitable for the high ionization [Fe~VI] zone. The derived
N$_e$ is somewhat lower than the N$_e$ range 2000 - 6300 cm$^{-3}$ 
obtained from several ionic spectra (including [O~II] and [S~II]) 
by \citet{hyu97}. Exactly similar results are obtained for 3 other line
ratios of [Fe~VI]; those have been omitted from Fig. 2 for brevity. 
The total observational uncertainties are 10\%,16\%,20\% and 23\% for
the 5485, 5336, 5177 and 5425 line ratios (with respect to the 5147 $\AA$
line), respectively (Hyung and Aller 1997).
However, an indication of the overall uncertainties
may be obtained from the 8th line ratio, 4973/5147,
which is independent of both T$_e$ and N$_e$ since both lines have the 
same upper level, and which therefore 
depends only on the ratio of the A-values; the observed value of 1.048
agrees closely with the theoretical value of 0.964.
Whereas the combined observational and theoretical uncertainties for 
any one line ratio can be significant, 
all measured line ratios yield a remarkably consistent N$_e$([Fe~VI]) and
substantiate the spectral model with FLE.

\section{TEMPERATURE-LUMINOSITY AND THE EMISSION REGION}
 
 Having determined the dependence of [Fe~VI] line intensities with FLE on local
quantities, the electron density and temperature in the
emission region \{N$_e$, T$_e$\} and related photon flux,
we next examine the dependence on the macroscopic parameters, the stellar
radiation temperature and the distance indicated by the dilution factor 
\{T$_{eff}$, W(r)\}. Fig. 3 shows the variation in two line ratios, with
W(r), at three T$_{eff}$ generally corresponding to the range of stellar
temperatures of the central stars of PN. It is seen that the correlation 
involving the photon density and the source radiation temperature is 
fairly well defined, analogous to the dependence on electron density N$_e$ in a 
characteristic range of electron temperatures T$_e$. 
It follows that
the $loci$ of the line ratios may be associated with photon and electron 
densities in the emission region and may be parametrized. A given range
of dilution factors, indicating the likely distances of the emission
region from the source, and the radiation intensity at the source, may
therefore be constrained to a most probable set of \{T$_{eff}$, W(r)\}.

 Fig. 4 illustrates a three-dimensional R(T$_{eff}$, W(r)) surface, intersected
by a plane defined by the observed line ratio R = R(obs)=0.460, along a contour
of \{T$_{eff}$,W(r)\}. The electron density and temperature are 
N$_e$ = 2000 cm$^{-3}$ and T$_e$ = 12,000 K, as derived from the 
line ratios in Fig. 2. It is seen that while the effective source
temperature and the distance of the emission region are not uniquely
defined as a 1-1 correspondence, they are constrained within a contour
by R(obs). In case of a source with a higher value of R(obs), the
contour \{T$_{eff}$,W(r)\} may be more stringently described.

\section{DISCUSSION AND CONCLUSION}

 Fluorescence in selected atomic species may be included in spectral
models to constrain the parameter space \{N$_e$,T$_e$,T$_{eff}$,W(r)\}
from theoretical and observed line ratios. 
The main parts of such an analysis are: (i) the
determination of  N$_e$ and T$_e$ from line ratios with and without FLE
(Fig. 2), (ii) examination of the luminosity function at T$_{eff}$ vs. W(r)
(Fig. 3), and (iii) calculation of possible R(T$_{eff}$,W(r)) contours
(Fig. 4). For sufficient precision it is necessary to study several line
ratios of the given ion, and to include all relevant atomic transitions in the
model. It is also required that the radiative and collisional atomic data be
know a $priori$ with high accuracy. Both of these criteria are
are now feasible with increasingly extensive and high-resolution
space and ground spectral observations, and large-scale theoretical
calculations such as under the Iron Project \citep{hum93}. 

 We have analyzed the spectra of a few other PN: NGC 6886, 
Hubble 12, NGC 2440 \citep{hyu95,hyu96,hyu98}. These
more detailed results will be presented in a companion publication 
\citep{che99c}. Interestingly, extragalactic spectral observations of 
[Fe~VI] lines at 5147 and 5177 $\AA$ have also been reported by \citet{mea91}
from PN SMP22 in the Small Magellanic Cloud. As with the galactic PN,
the observed 5177/5147 line ratio of 0.83 in SMC SMP22 does not correspond
to realistic (N$_e$,T$_e$) without FLE. An optimum parameter fit with 
FLE however yields N$_e$([Fe~VI]) $\approx$ 1000 cm$^{-3}$, at T$_e$ = 20,000 K
(Meatheringham and Dopita report T$_e$([O~III]) = 26,600 K), and
W(r) = 10$^{-13}$, T$_{eff}$ = 150,000 K. Strictly speaking W(r)
$\longrightarrow$ R$_{\ast}$/r; i.e. it yields the ratio of the stellar
radius to the distance of the emission region. 
It is also
important to note that the determination of total and fractional 
element abundances could be highly uncertain if the effect of FLE on
spectral formation is not ascertained accurately.

 The proposed approach could possibly be employed to study
variable central luminosity sources such as novae and AGN. With variations in
electron density and temperature, and the spatial extent of the emission
region, a given line ratio  may exhibit large
variations with temperature-luminosity, as illustrated in Figs. 3 and 4.
It may also be noted that the method is not predicated on the assumption
of coronal or photoionization equilibrium, generally assumed in modeling
nebulae and AGN. However,
further constraints may be sought from calculations such as
coronal, and/or photoionization, models in ionization equilibrium that
determine the ionization fraction of an element vs. r. Conversely,
determination of peak abundance fractions from models may be verified
through direct spectroscopic analysis of FLE dominated emission from an
ion, as described herein. For objects where the nebulosity is 
observed and spatially defined, determination of T$_{eff}$
(temperature-luminosity) of the source may yield
independent distance estimates via triangulation.

In non-ionization equilibrium the
spectral modeling might entail
level-specific ionization~/~recombination involving metastable
populations (work in progress).
Non-thermal specific luminosity
or a mean photon intensity function vs. frequency 
should be applied in real applications, e.g. in the synchrotron continuum
pumping in Crab nebula (Davidson \& Fesen 1985; Lucy 1995), or a
power-law spectrum in AGN.
Future FLE work in intense radiation background sources involves the analysis
of the symbiotic nova RR Tel \citep{mck97} and the star V 1016 Cygni (a
possible single PN in formation or a symbiotic nova)
 \citep{ahe77}, where
many [Fe~VI] emission lines have been observed, the spectra
of `coronal' iron ions Fe~XVIII - XVII in narrow
or coronal line region (NLR,CLR) of AGN, and [Ni~II] and [Co~II]
lines in nebular remnants of supernovae type II (e.g. 1987A) and Ia. 

 Given that spectral formation from an atomic species is localized 
with respect to the radiative source, and subject to FLE, variation in spectral 
intensities with photon density should enable
qualitative and quantitative constraints on the nature of the source
and the emission region. To that end,
the radiation field could be thermal or non-thermal, and the geometrical
dilution with distance to the emission region could be suitably parameterized, 
including local variables such as electron density and temperature
(radiative transfer effects would be important at significant optical
depths).  

 Finally we note that in addition to demonstration of (a) FLE, and (b)
regular intensity variations in line emitting regions, studies of (a)
and (b) may resolve discrepancies in element
abundance determinations from recombination lines and using 
collisionally excited lines without FLE (Seaton 1978).

\acknowledgments
 This work was partially supported by the NSF (AST-9870089) and NASA
(NAG5-8423).


\def\r{\leftskip10pt \parindent-10pt \parskip0pt}
\def\apj{ApJ}
\def\apjs{ApJS}
\def\apjl{ApJL}
\def\aj{Astron. J}
\def\pasp{Pub. Astron. Soc. Pacific}
\def\mn{MNRAS}
\def\aa{A\&A}
\def\aasup{A\&A Suppl.}
\def\baas{Bull. Amer. Astron. Soc.}
\def\jqsrt{J. Quant. Spectrosc. Radiat. Transfer}
\def\jpb{Journal of Physics B}
\def\pra{Physical Review A}
\def\adndt{Atomic Data And Nuclear Data Tables}

\newpage
\centerline{\bf Figure Captions}

\figcaption[fig1.eps]{Schematic diagram of fluorescent excitation (FLE)
of [Fe~VI] optical emission lines. The full radiative-collisional model
includes 80 fine structure levels. 
\label{fig1}}

\figcaption[fig2.eps]{Diagnostic line ratios for PN NGC 6741
\citet{hyu97}, with and without FLE of Fe~VI. The W = 0 (no
FLE) does not correspond to any observed line ratio.
\label{fig2}}

\figcaption[fig3.eps]{Variation of line ratio with geometrical dilution factor
W(r),  and the effective radiation temperature T$_{eff}$, 
indicating the distances and localization of the emission region.
\label{fig3}}

\figcaption[fig4.eps]{Spectral intensity variation vs. \{T$_{eff}$, W(r)\}. 
Intersection of the plane defined by the observed line ratio
in NGC 6741 with this surface describes a contour of possible set of \{T$_{eff}$, W(r)\}. The
electron density and temperature \{N$_e$, T$_e$\} are determined as in Fig. 2. \label{fig4}}

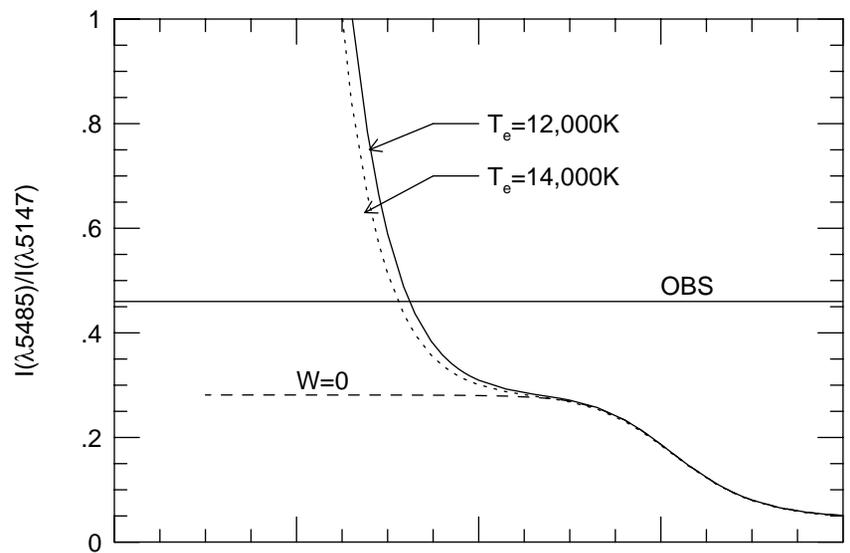
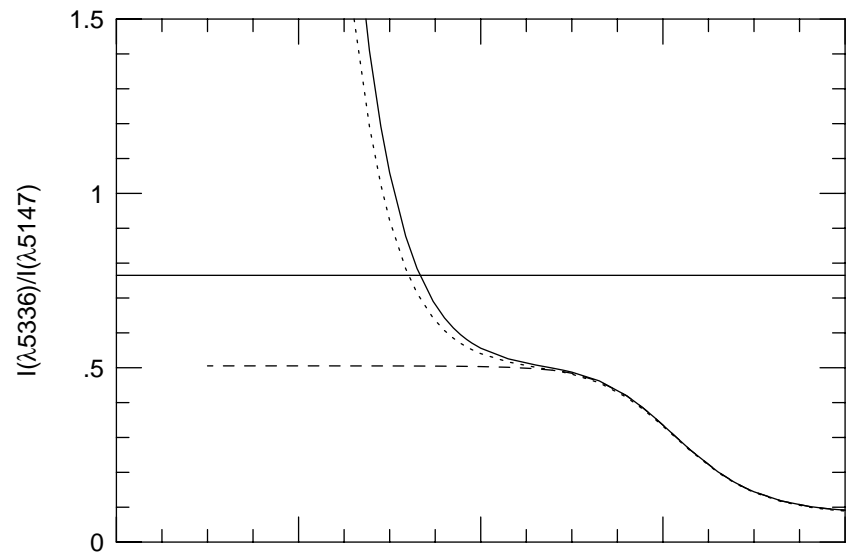
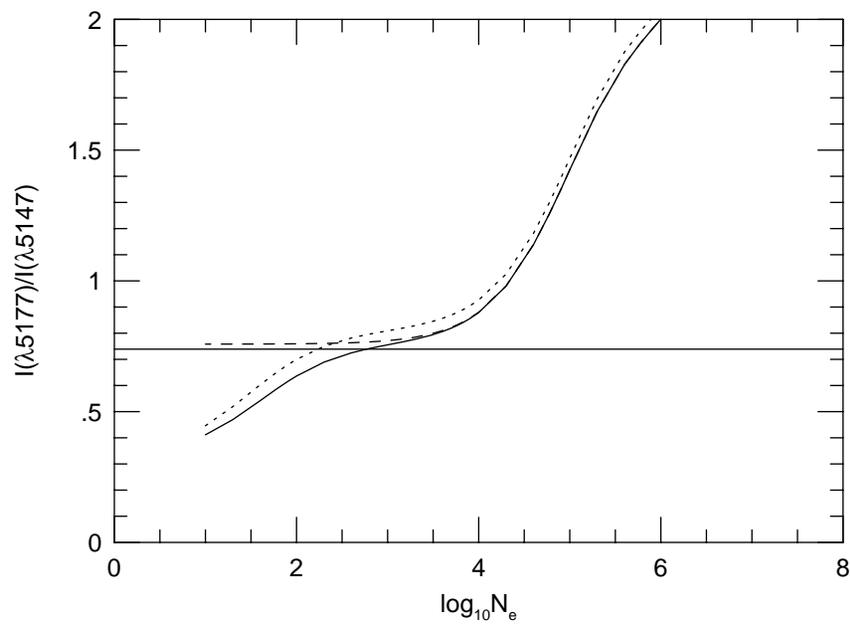
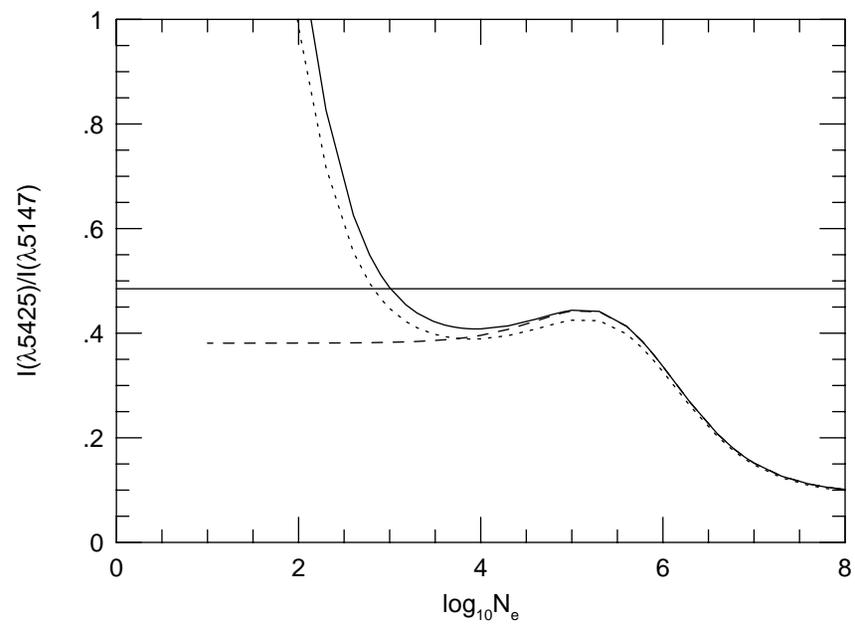
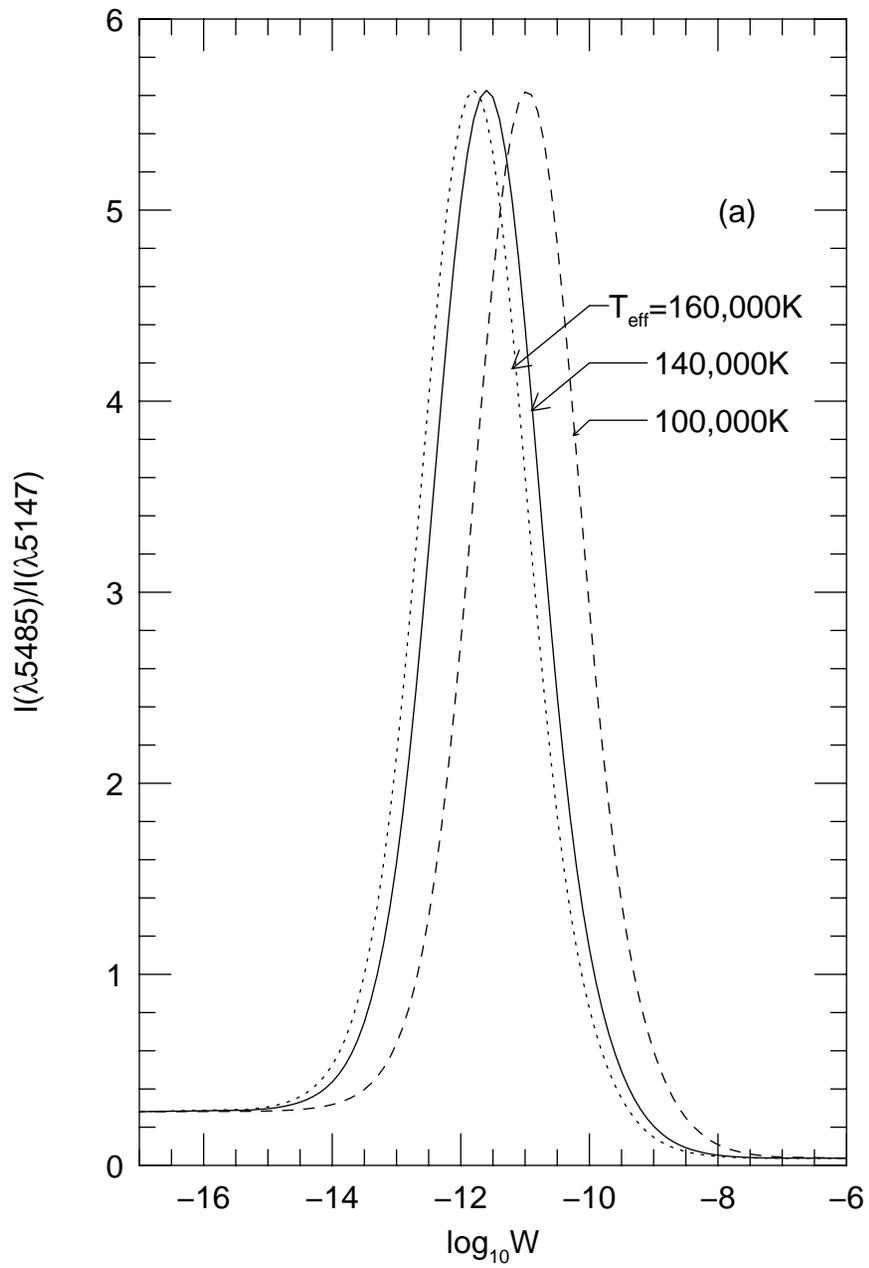
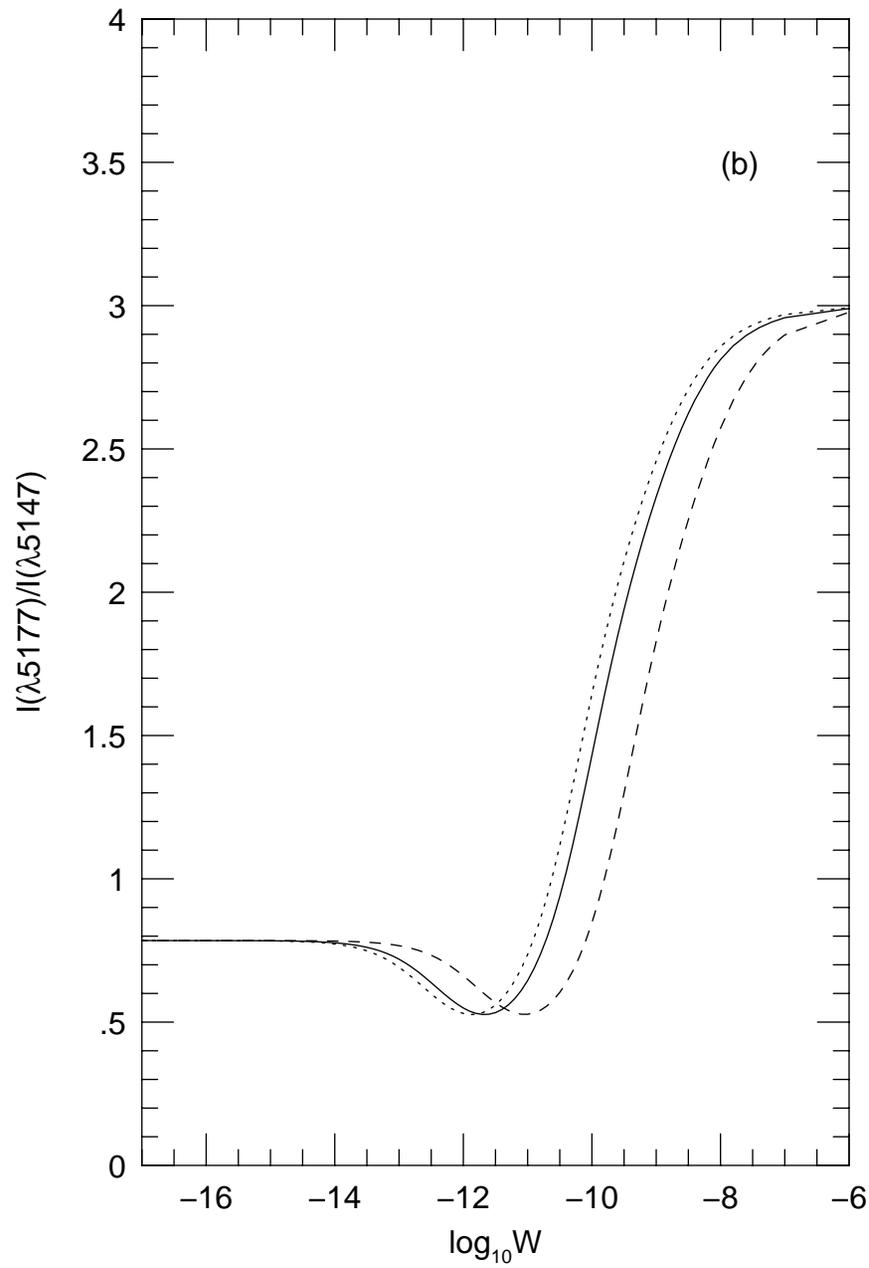
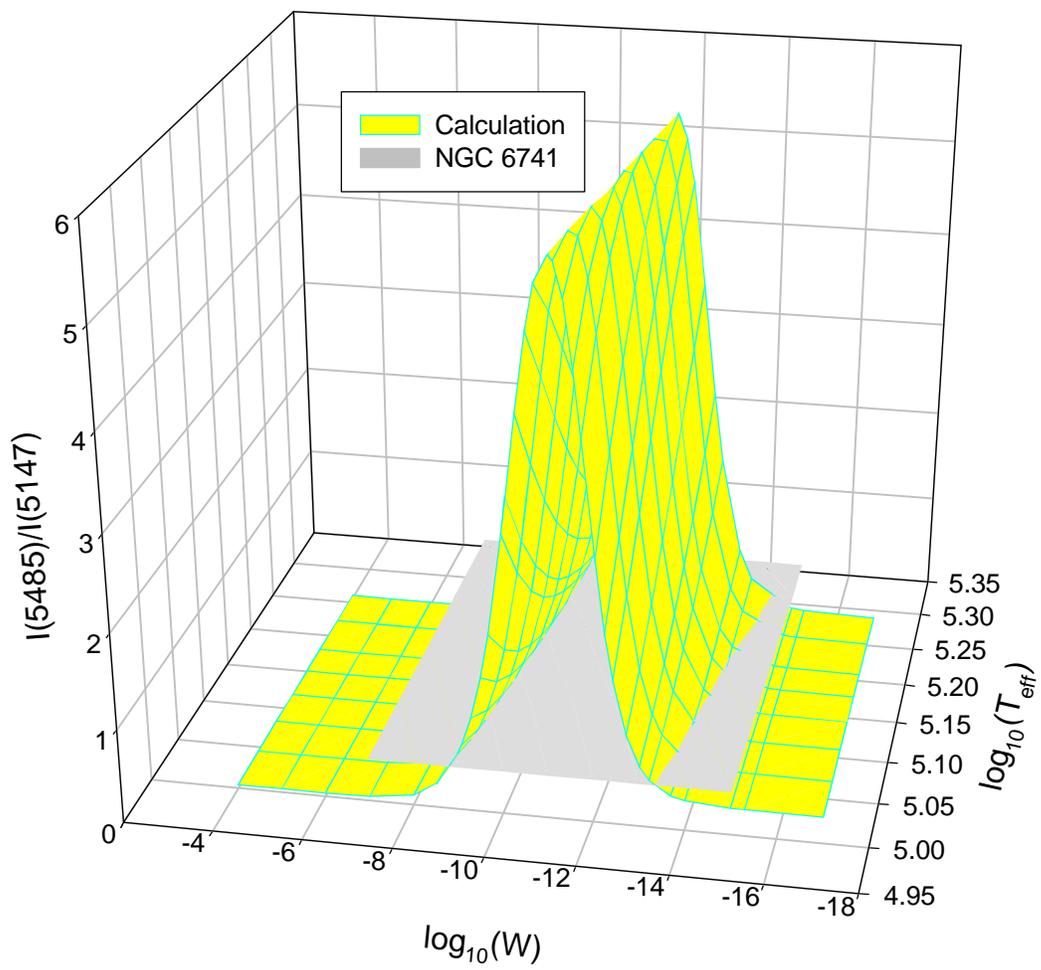
\end{document}